\documentclass[11pt, a4paper]{article}
\usepackage{}

\usepackage{indentfirst}
\usepackage{lineno}
\usepackage{graphicx}
\usepackage{ae}
\usepackage{amsmath}
\usepackage{amssymb}
\usepackage{latexsym}
\usepackage{url}
\usepackage{epsfig}
\usepackage{cite}
\usepackage{mathrsfs}
\usepackage{amsfonts}
\usepackage{amsthm}
\usepackage{float}
\usepackage{booktabs}
\usepackage{subfigure}
\usepackage{multirow}

\usepackage{color}

\long\def\delete#1{}

\newcommand{\be}{\begin{equation}}
\newcommand{\ee}{\end{equation}}
\newcommand{\ben}{\begin{equation*}}
\newcommand{\een}{\end{equation*}}
\newcommand{\bea}{\begin{eqnarray}}
\newcommand{\eea}{\end{eqnarray}}
\newcommand{\bean}{\begin{eqnarray*}}
\newcommand{\eean}{\end{eqnarray*}}

\numberwithin{equation}{section}

\title{Two new methods for identifying proteins based on the domain protein complexes and topological properties\thanks{Supported by the National Natural Science Foundation of China (No.11361033) and the Natural Science Foundation of Gansu Province (No.1212RJZA029).}}

\author{Pengli Lu\thanks{Corresponding author.   E-mail
addresses: lupengli88@163.com (\textbf{P. Lu}), yujingjuanmercy@163.com (\textbf{J. Yu}).} \;and\; JingJuan Yu
\\
\footnotesize{School of Computer and Communication, Lanzhou University of Technology, Lanzhou, 730050, Gansu, P.R. China}}

\date{}

\begin{document}

\openup 0.5\jot
\maketitle

\begin{abstract}
The recognition of essential proteins not only can help to understand the mechanism of cell operation, but also help to study the mechanism of biological evolution. At present, many scholars have been discovering essential proteins according to the topological structure of protein network and complexes. While some proteins still can not be recognized. In this paper, we proposed two new methods complex degree centrality ($CDC$) and complex in-degree and betweenness definition ($CIBD$) which integrate the local character of protein complexes and topological properties to determine the essentiality of proteins. First, we give the definitions of complex average centrality ($CAC$) and complex hybrid centrality ($CHC$) which both describe the properties of protein complexes. Then we propose these new methods $CDC$ and $CIBD$ based on $CAC$ and $CHC$ definitions. In order to access these two methods, different Protein-Protein Interaction (PPI) networks of Saccharomyces cerevisiae, DIP, MIPS and YMBD are used as experimental materials. Experimental results in networks show that the methods of $CDC$ and $CIBD$ can help to improve the precision of predicting essential proteins.
\bigskip

\bigskip

\end{abstract}

\section{Introduction}
Protein is one of the main components of human life. Essential protein is defined as a protein which would result in the inability of the organism to survive when it is removed by a knockout mutation. Essential proteins are more conserved in biological evolution in comparision to non-essential proteins [1]. Not only can essential proteins help us understand the growth control system of cells, and then understand the mechanism of life, but also help the study of biological evolution mechanism [2]. Removing essential proteins can lead to fatal or infertility [3]. Determining the essentiality of proteins is of great significance to the research of system biology which provides valuable theories and methods for the diagnosis of diseases, drug design, etc. [4]. Therefore, identifying the essential protein is meaningful in biomedicine.

Previous methods for identifying essential proteins mainly used some biological experiments, including conditional knockouts [5], RNA interference [6], and single gene knockouts [7], coupled with the survival ability of infected organisms being tested. However, these biological experimental processes not only consume amounts of time and costs, but also require a lot of biological resources. Nowadays, it has been a crucial research direction in the field of bioinformatics for predicting essential proteins from a large number of biological experiments by using computer technology theory and research methods.

Jeong H M et al. put forward that the essentiality of proteins is associated with the topological structure in protein interaction networks [8]. There are some species including S.cerevisiae, E.coli, C.elegans and D.melanogaster that have demonstrated the hubs in PPIs have more chance to be essential proteins [9]. Thus, we are working to investigate the importance of proteins in topologies to essential proteins. On the basis of network topology characteristics of nodes, there are many centrality measures to discover essential proteins. Some of them are global network characteristics, like betweenness centrality ($BC$) [11,38], eigenvector centrality ($EC$) [19], information centrality ($IC$) [20] and closeness centrality ($CC$) [13]. Others are local network features, such as degree centrality ($DC$) [10,14,15], subgraph centrality ($SC$) [16], local average centrality ($LAC$) [17] and topology potential-based method ($TP$) [34]. On the basis of network topology characteristics of edges, there are also some measures, including edge clustering coefficient ($ECC$) [35], and improved node and edge clustering coefficient ($INEC$) [36]. In recent years, many scholars have been working to identify proteins in combination with protein information, such as $PeC$ which combines edge clustering coefficients with gene expression data correlation coefficients [24], $esPOS$ which using gene expression information and subcellular localization information [21], $SPP$ which based on sub-network partition and prioritization by integrating subcellular localization [12], extended pareto optimality consensus model ($EPOC$) that fuses neighborhood closeness centrality and Orthology information [39]. Go terms information can also be used to predict essential proteins such as $RSG$ method in [25].

Apart from analyzing the essentiality of proteins from topological point of view and protein information, analyzing the characteristics from the perspective of protein complexes has become another direction of our study. Hart G T et al. found that the essential proteins are often determined by the protein complexes in which the protein is involved, rather than by a single protein [22]. Li et al. also prove that the frequency of the essential proteins appear in the complex would be more than that in the whole network [21,41]. To give examples, Luo J W et al. raised the local interaction density of binding protein complexes ($LIDC$) for predicting essential proteins [37]. Qin C et al. put forward the $LBCC$, a measure on the basis of both network topology features and protein complexes [18]. Li et al. proposed united complex centrality ($UC$) which combine the edge clustering coefficient and the freqencies of proteins appeared in complexes [23]. From the results of their experiences, we can see that the performances of these methods are better than using the pure topological methods.

Therefore, on the basis of the association with protein complexes information and topological properties, our two new novel methods complex degree centrality ($CDC$) and complex in-degree and betweenness definition ($CIBD$) are proposed. In order to describe the structural properties of protein complexes, we define $CAC$ and $CHC$ of a node $v$. Between the two indicators we put forward, one is called $CDC$ which combine the node and its neighbors properties to describe the features for protein complexes, the other is called $CIBD$ based on the features of protein complexes, local features and global properties in the network.

To assess the quality of $CDC$ and $CIBD$ methods, we apply them to different datasets of Saccharomyes cerevisiae, DIP, MIPS and YMBD. In order to obtain the performance of our proposed methods, we make comparisions by using some existing measures, including $DC$, $BC$, $LAC$, $SC$, $LBCC$, $EC$, $SoECC$ and $UC$ which can gain the original paper from [10], [11], [17], [16], [18] ,[19], [28] and [23] respectively. In terms of the sensitivity, specificity, positive predictive value, negative predict value, F-measure, accuracy rate and the evaluation methods of ``sorting-screening", the precision-recall curves and jackknife, the results show that our two methods are more effective in determining the essentiality of proteins than existing measures.


\section{Methods}
\subsection{Notation}
An undirected simple graph $G(V, E)$ can be used to express a  network of protein interaction. Proteins can be regarded as nodes set $V$ of a network and the connections between two proteins can be regarded as edges set $E$. The number of nodes and edges in a graph $G$ can be defined as $|V(G)|$ and $|E(G)|$ separately. The neighbor set of node $v$ is denoted by $N_{v}$, and its number can be represented as $|N_{v}|$. The induced subgraph of $G[S]$ is a subgraph of $G$ induced by the nodes set $S$.

\subsection{Previously Proposed Centrality Measures}
There are some centralities we need to understand.
\begin{itemize}
\item Betweenness centrality ($BC$) [11]
    \begin{equation}\label{adjmatix}
    \begin{aligned}
    \begin{split}
    BC(v)=\displaystyle\sum_{s \neq v \neq t \in V} \frac{\sigma_{st}(v)}{\sigma_{st}}
    \end{split}
    \end{aligned}
    \end{equation}
    where $\sigma_{st}$ denotes the number of shortest paths between $s$ and $t$. $\sigma_{st}(v)$ denotes the number of shortest paths from $s$ to $t$ that pass through the node $v$.
\item In-degree centrality of complex ($IDC$) [21]
    \begin{equation}\label{adjmatix}
    \begin{aligned}
    \begin{split}
    IDC(v)=\sum_{i\in ComplexSet(v)} IN-Degree(v)_{i}
    \end{split}
    \end{aligned}
    \end{equation}
   A subset of protein complexes that containing protein $v$ can be represented as $ComplexSet(v)$, the degree of node $v$ for the $i_{th}$ protein complex which belongs to $ComplexSet(v)$ can be represented as $IN-Degree(v)_{i}$.

\item $LBCC$ method [18]
    \begin{equation}\label{adjmatix}
    \begin{aligned}
    LBCC(v)=a*\log Den_{1}(v)+b*\log Den_{2}(v)
    \\+c*\log IDC(v)+d*\log BC(v)
    \end{aligned}
    \end{equation}
    Specifically,
    \begin{equation}\label{adjmatix}
    \begin{aligned}
    \begin{split}
    Den_{1}(v)=\frac{2|E(H)|}{|V(H)|(|V(H)|-1)}
    \end{split}
    \end{aligned}
    \end{equation}
where the induced subgraph $G[N_{v}\bigcup \{v\}]$ can be represented as $H$.
    \begin{equation}\label{adjmatix}
    \begin{aligned}
    \begin{split}
    Den_{2}(v)=\frac{2|E(H)|}{|V(H)|(|V(H)-1)}
    \end{split}
    \end{aligned}
    \end{equation}
    where $M_{u}=\bigcup _{u\in N_{v}}N_{u}$, $H$ represents the induced subgraph $G[M_{u}\bigcup N_{v}\bigcup\{v\}]$.
\end{itemize}

\subsection{New Centrality: $CDC$ and $CIBD$}

The basic considerations of $CDC$ and $CIBD$ are as follows: (1)The essential proteins appear in complexes can be more frequency. (2)Both the node itself and its neighbors are critical to affect the essentiality. (3)The global topological is considered to be a factor in locating essential proteins. Consequently, we present two new definitions to judge the essentiality of proteins by combining the domain features of protein complex and the topological properties.

First, we present a new complex average central definition ($CAC$) for the neighbors of a node $v$,
    \begin{equation}\label{adjmatix}
    \begin{aligned}
    \begin{split}
    CAC(v)=\frac{\sum_{u\in N_{v}}IDC(u)}{|N_{v}|}
    \end{split}
    \end{aligned}
    \end{equation}
    where $\sum_{u\in N_{v}}IDC(u)$ represents the total values of $IDC$ for all the neighbors of a node $v$. $IDC$ centrality has been mentioned in Eq. (2)

Then, we propose complex hybrid central definition ($CHC$) by combining the number of complexes for a node $v$ with complex average central definition $CAC$,
 \begin{equation}\label{adjmatix}
    \begin{aligned}
    \begin{split}
    CHC(v)=N_{complex}(v)\cdot CAC(v)\cdot IDC^{2}(v)
    \end{split}
    \end{aligned}
    \end{equation}
    where $N_{complex}(v)$ denotes the total number of complexes for a node $v$.

Now, based on the two definitions that we described above, we propose these two new methods for estimating the essentiality of a node $v$. One is complex degree centrality ($CDC$) which combine the node with its neighbors to describe the properties for protein complexes,
    \begin{equation}\label{adjmatix}
    \begin{aligned}
    \begin{split}
    CDC(v)=a*CAC(v)+b*IDC(v)
    \end{split}
    \end{aligned}
    \end{equation}
where $a$, $b$ are random parameters ranging from $1$ to $10$. After conducting plenty of experiments, we can get the best results of the method $CDC$ when $a$ and $b$ are $1$ and $4$, respectively.

The other is complex in-degree and betweenness definition ($CIBD$) which combining $CHC$, $Den_{2}$ and $BC$, where the structural property of the protein complexes is described by $CHC$, the local feature is described by $Den_{2}$ and the global property is described by $BC$. Since the values of these measures are quite different, the data is normalized by logarithmic transformation,
\begin{equation}\label{adjmatix}
    \begin{aligned}
    \begin{split}
    CIBD(v)=a*\log (CHC(v))+b*\log (Den_{2})
    \\+c*\log (BC(v))
    \end{split}
    \end{aligned}
    \end{equation}
where $a$, $b$ and $c$ are random parameters ranging from $1$ to $10$. Under the amounts of experiments, we can get the best results of the method $CIBD$ when $a$, $b$ and $c$ are $1$, $3$ and $1$, respectively.

The descirption of $CDC$ and $CIBD$ algorithms are in Table 1.

\begin{table}[!htbp]
\centering
\scriptsize
\caption{Description~of~CDC~and~CIBD~algorithms}
\setlength{\tabcolsep}{10pt}
\begin{tabular}{l}
 \toprule
 $CDC$ and $CIBD$ algorithms\\
 \midrule
 $\mathbf{Input:}$ Undirected graph $G=(V(G),E(G))$ stands for \\
 a PPI network, $C=\{C_{i}=(V(C_{i}),E(C_{i}))|C_{i}\subset G \}$\\ represents complexes\\
 $\mathbf{Output:}$ The proteins list sorted by $CDC$, $CIBD$ in a \\descending order\\
 $\mathbf{01:}$ $\mathbf{For}$ each vertex $v\in V(G)$ $\mathbf{do}$ $IDC(v)=0$\\
 $\mathbf{02:}$ ~~~~$\mathbf{For}$ each $\forall C_{i}\in C$ $\mathbf{do}$\\
 $\mathbf{03:}$ ~~~~~~~~calculate $IDC(v)=IDC(v)+IN-Degree(v)_{i}$\\
 //where $IN-Degree(v)_{i}$ is the value of $DC(v)$ in $i_{th}$ complex\\
 $\mathbf{04:}$ $\mathbf{For}$ each vertex $v\in V(G)$ $\mathbf{do}$\\
 $\mathbf{05:}$ ~~~~Find the neighbor nodes $N_{v_{1}}$ of node $v$\\
 //where $N_{v_{1}}$ stands for the neighbor nodes set for node $v$\\
 $\mathbf{06:}$ ~~~~calculate $CAC(v)$ by Equation(6)\\
 $\mathbf{07:}$ ~~~~$\mathbf{For}$ each vertex $v_{2}\in N_{v_{1}}$ $\mathbf{do}$\\
 $\mathbf{08:}$ ~~~~~~~~Find the neighbor nodes of $N_{v_{2}}$ \\
 //where $N_{v_{2}}$ stands for the neighbor nodes set for node $v_{2}$ \\
 which $v_{2}\in N_{v_{1}}$\\
 $\mathbf{09:}$ ~~~~~~~~calculate $Den_{2}$ by Equation(5)\\
 $\mathbf{10:}$ $\mathbf{For}$ each vertex $v\in V(G)$ $\mathbf{do}$ \\
 $\mathbf{11:}$ ~~~~calculate $CHC(v)$ by Equation(7)\\
 $\mathbf{12:}$ calculate and sort $CDC(v)$ by Equation(8) \\
 $\mathbf{13:}$ calculate and sort $CIBD(v)$ by Equation(9) \\
 \bottomrule
\end{tabular}
\end{table}

\section{Experimental data and assessment methods}
\subsection{Experimental data}

In order to analyze the performance of these two algorithms of $CDC$ and $CIBD$, experiments are conducted by using the protein interaction data of Saccharomyes cerevisiae because its proteins are more complete.

Three sets of PPI network data YDIP, YMIPS and YMBD are used. The DIP dataset is marked as YDIP network [26]; The MIPS dataset is marked as YMIPS network [25]; The YMBD network comes from the Mark Gerstein Lab website. In the protein network, all self-interaction and repetitive interaction are deleted as a data preprocessing of these PPIs.
Specific properties for these three networks are presented in the Table 2. In the YDIP network, there are 5093 proteins and 24743 interactions, whose clustering coefficient is about 0.0973. YMIPS network includes 4546 proteins and 12319 interactions, whose clustering coefficient is about 0.0879. YMBD network includes 2559 proteins and 11835 interactions, whose clustering coefficient is about 0.4445.

The known essential protein is derived from four databases: MIPS [40], SGD (Saccharomyces Genome Database) [33], SGDP (Saccharomyces Genome Deletion Project) [4], and DEG (Database of Essential Genes) [27]. The protein complex set is from CM270 [40], CM425 [29], CYC408 and CYC428 datasets [30,31] which can gained from [21], containing 745 protein complexes (including 2167 proteins).

\begin{table}[!htbp]
\centering
\caption{Data details of the three protein networks: YDIP, YMIPS, YMBD}
\setlength{\tabcolsep}{4pt}
\begin{tabular}{|c|c|c|c|c|c|c|}
\hline
Dataset & Proteins & Interactions & Average degree & Essential proteins &Clustering coefficient\\\hline
YDIP & 5093 & 24743 & 9.72 & 1167 & 0.0973  \\\hline
YMIPS & 4546 & 12319 & 5.42 & 1016 & 0.0879  \\\hline
YMBD & 2559 & 11835 & 9.25 & 763 & 0.4445 \\
\hline
\end{tabular}
\end{table}

\subsection{Assessment methods}

According to their values of $CDC$, $CIBD$ and other eight prediction measures including $DC$, $BC$, $EC$, $SC$, $LAC$, $LBCC$, $SoECC$ and $UC$, proteins are sorted from high to low orders. First, we choose some number of top proteins in sequence as predictive essential proteins and then compare them with the real essential proteins. This allows us to know the quantity of true essential proteins. Therefore, the sensitivity ($SN$), specificity ($SP$), F-measure ($F$), accuracy ($ACC$), positive predictive value ($PPV$) and negative predictive value ($NPV$) can be calculated [28,29].

The following are the formulas for calculating these six statistical indicators.

Sensitivity: $$SN=\frac{TP}{TP+FN}$$

Specificity:
$$SP=\frac{TN}{TN+FP}$$

Positive predictive value:
$$PPV=\frac{TP}{TP+FP}$$

Negative predictive value:
$$NPV=\frac{TN}{TN+FN}$$

F-measure:
$$F=\frac{2*SN*PPV}{SN+PPV}$$

Accuracy:
$$ACC=\frac{TP+TN}{P+N}$$
where $TP$ stands for the number of true essential proteins which are correctly selected as essential proteins. $FP$ is the number of nonessential proteins which are incorrectly selected as essential. $TN$ is the number of nonessential proteins which are correctly selected as nonessential. $FN$ is the number of essential proteins which are incorrectly selected as nonessential. $P$ and $N$ stand for the sum number of essential and nonessential proteins, respectively.


\section{Results}
\subsection{Comparison with other previously proposed measures}

\begin{figure}[htbp]
\centering
\includegraphics[width=6cm]{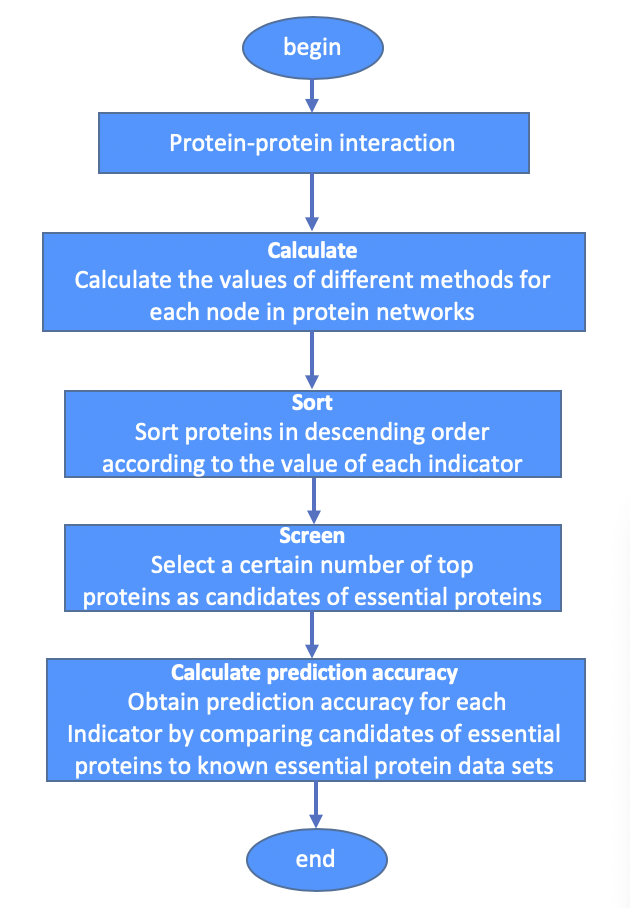}
\scriptsize\\
{Fig. 1~``sorting-screening" method}
\end{figure}

\begin{figure}[htbp]
\centering
\includegraphics[width=3in]{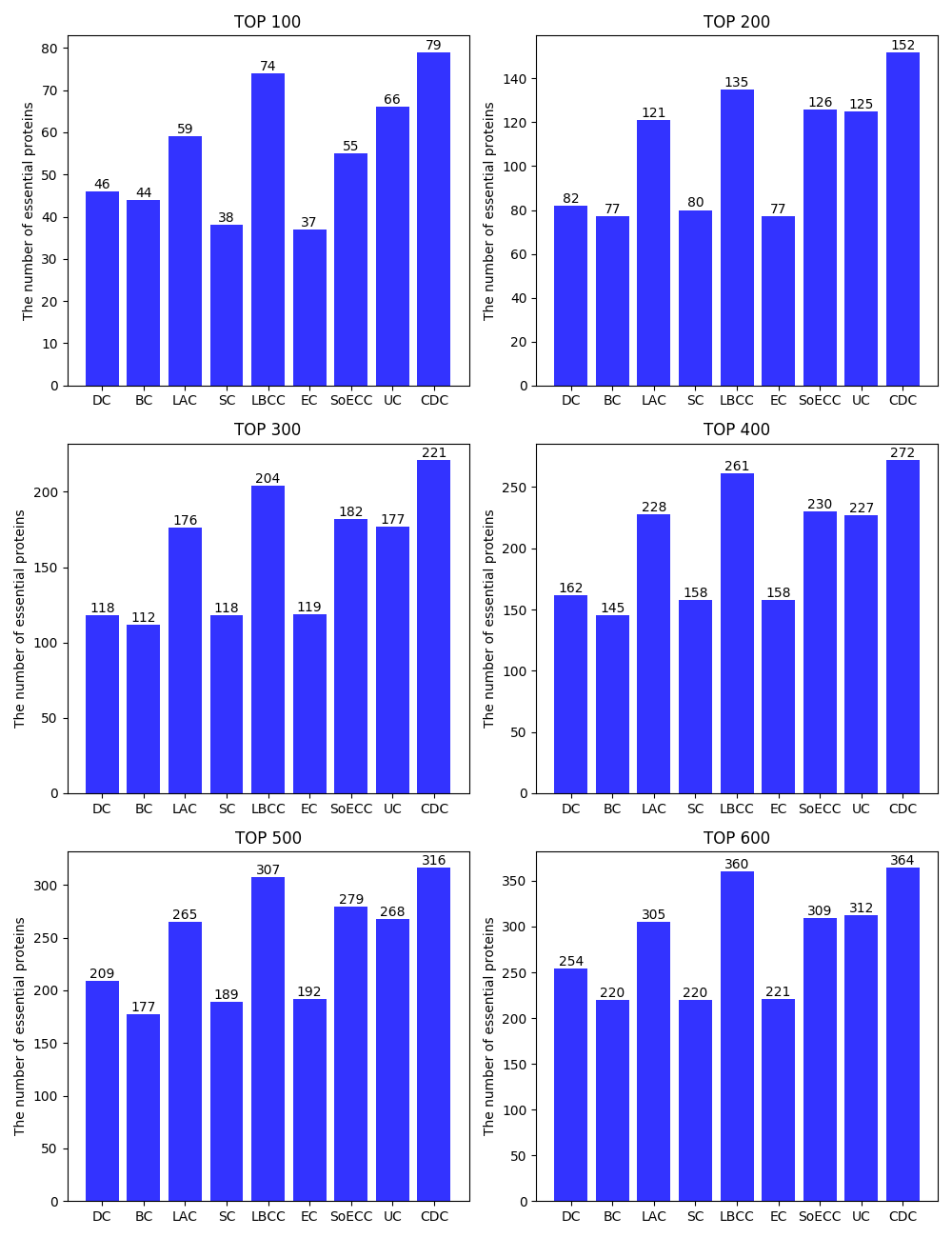}
\scriptsize\\{Fig. 2~The quantity of true essential proteins determined by $CDC$ and other eight previously methods from the YDIP network.}
\end{figure}

\begin{figure}[htbp]
\centering
\includegraphics[width=3in]{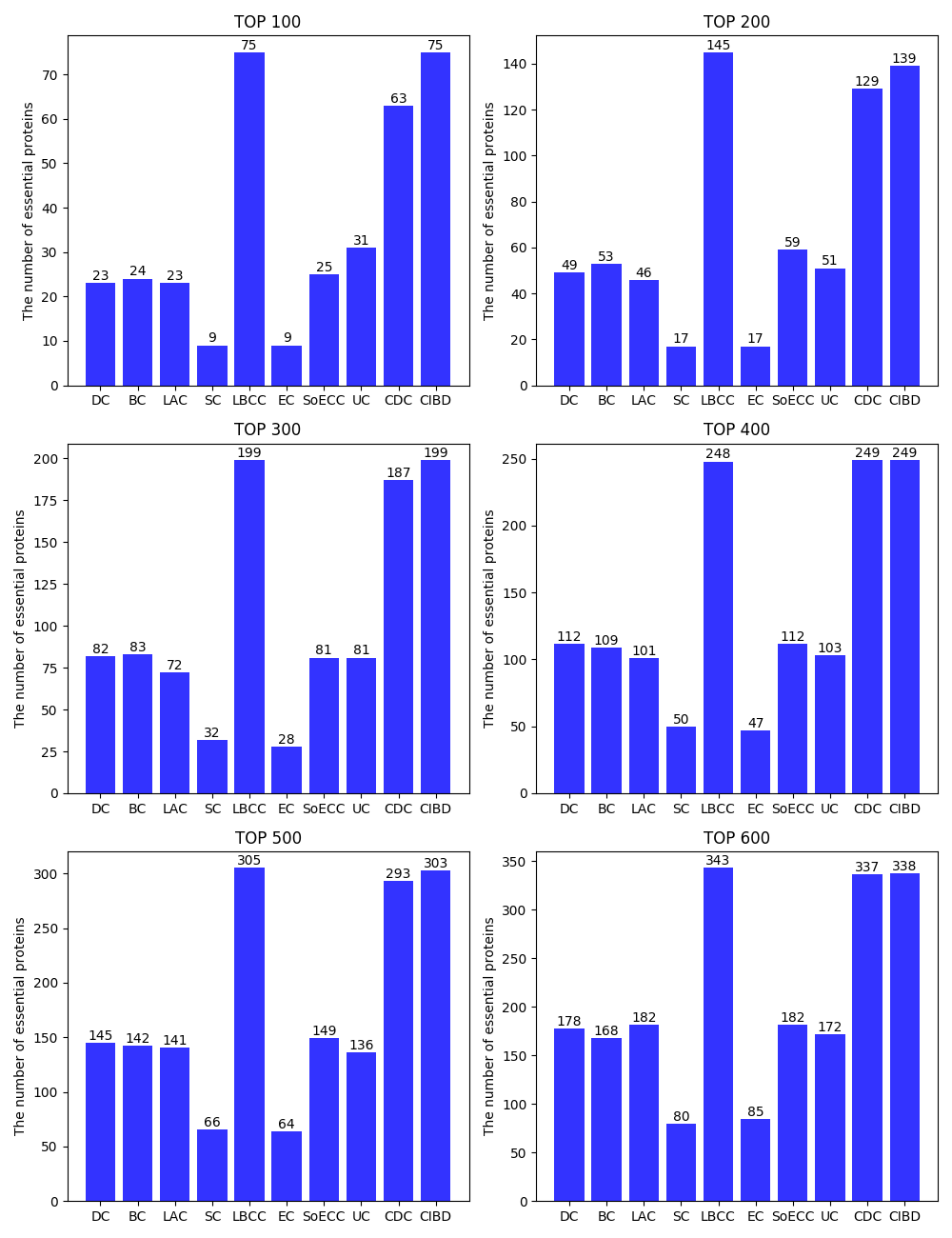}
\scriptsize\\{Fig. 3~The quantity of true essential proteins determined by $CDC$, $CIBD$ and other eight previously methods from the YMIPS network.}
\centering
\end{figure}

\begin{figure}[htbp]
\centering
\includegraphics[width=3in]{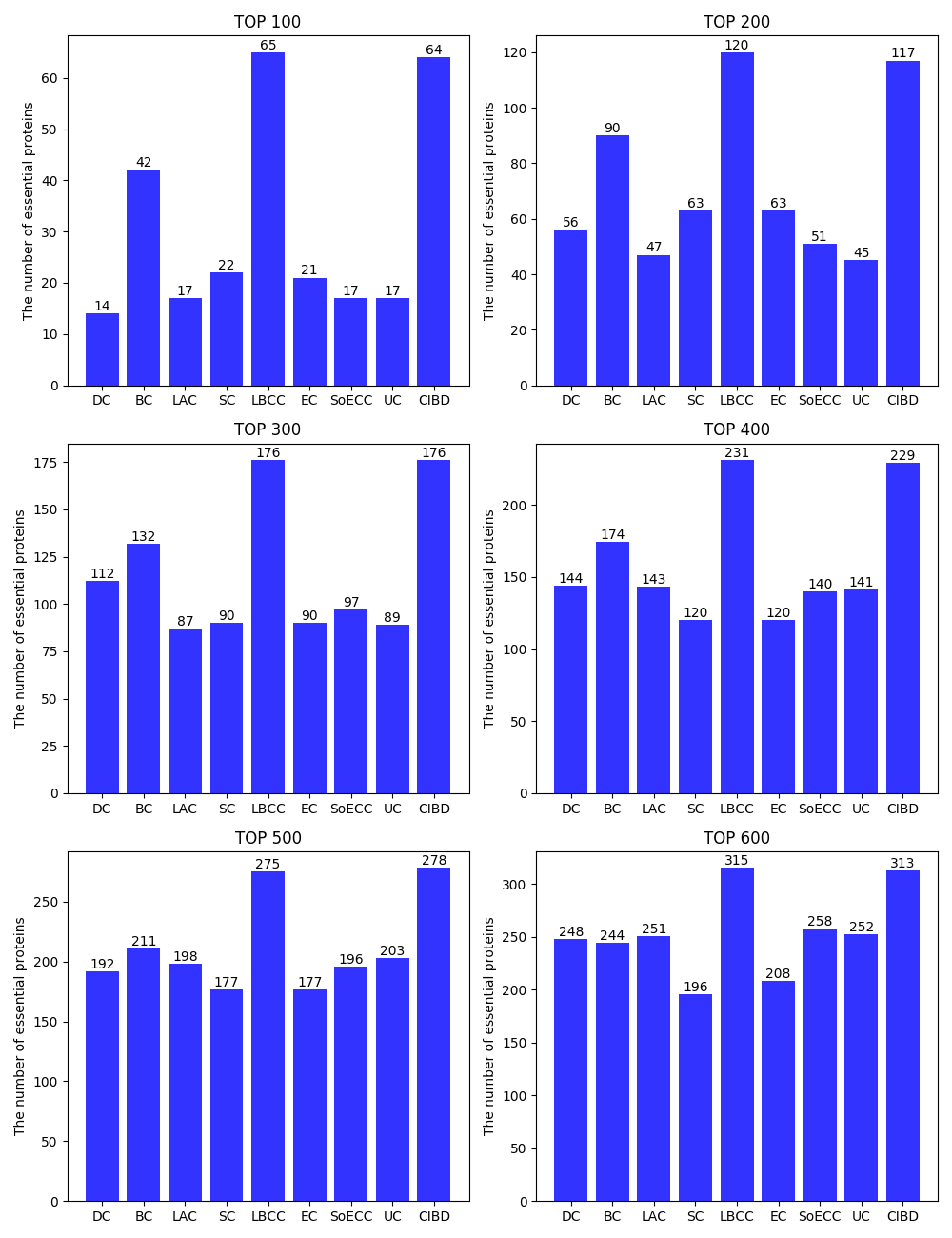}
\scriptsize\\{Fig. 4~The quantity of true essential proteins determined by $CIBD$ and other eight previously methods from the YMBD network.}
\centering
\end{figure}

In this paper, to evaluate the efficiency and accuracy of different indicators in identifying essential proteins, we follow the principle of ``sorting-screening" which has described as a flow chart in Fig. 1. Then we compare $CDC$ and $CIBD$ methods with other eight previous measures including $DC$, $BC$, $EC$, $SC$, $LAC$, $LBCC$, $SoECC$ and $UC$ in the three datasets. The algorithm for $LBCC$ was implemented according to [18] which used the same datasets as ours. Other algorithms of $DC$, $BC$, $EC$, $SC$, $LAC$, $SoECC$ and $UC$ were implemented according to references [10], [11], [19], [16], [17], [28] and [23] respectively. Besides, we can also get these algorithms by using CytoNCA [42], which is a Cytoscape app for network centrality. We have mentioned the method of $BC$ and $LBCC$ in the Section Previously Proposed Centrality Measures. Now we give a brief description of other six indicators.

\begin{itemize}
\item Degree centrality ($DC$) [10]
    \begin{equation}\label{adjmatix}
    \begin{aligned}
    \begin{split}
    DC(v)= deg(v)
    \end{split}
    \end{aligned}
    \end{equation}
    where $deg(v)$ denotes the degree of a node $v$.

\item Local average connectivity centrality ($LAC$) [17]
    \begin{equation}\label{adjmatix}
    \begin{aligned}
    \begin{split}
    LAC(v)=\frac{\sum_{u \in N_{v}} deg^{C_{v}} (u) } {|N_{v}|}
    \end{split}
    \end{aligned}
    \end{equation}
 where $C_{v}$ is the subgraph induced by the node set $N_{v}$ of $G$ and $deg^{C_{v}}(u)$ is the number of its neighbors in $C_{v}$ for a node $u\in Nv$.

\item Subgraph centrality ($SC$) [16]
    \begin{equation}\label{adjmatix}
    \begin{aligned}
    \begin{split}
    SC(v)=\sum_{k=0}^{\infty}\frac{\mu_{k}(v)}{k!}
    \end{split}
    \end{aligned}
    \end{equation}
 where $\mu_{k}(v)$ denotes the number of closed walks of length $k$ which starts and ends at node $v$.

\item Eigenvector centrality ($EC$) [19]
    \begin{equation}\label{adjmatix}
    \begin{aligned}
    \begin{split}
    EC(v)=\alpha_{max}(v)
    \end{split}
    \end{aligned}
    \end{equation}
   where $\alpha_{max}$ refers to the main eigenvector corresponding to the largest eigenvalue of the network adjacency matrix $A$, and $\alpha_{max}(v)$ represents the $v_{th}$ component of $\alpha_{max}$.

\item The sum of edge clustering coefficients ($SoECC$) [28]
    \begin{equation}\label{adjmatix}
    \begin{aligned}
    \begin{split}
    ECC_{v,u}=\frac{z_{v,u}}{min(k_{v}-1,k_{u}-1)}
    \end{split}
    \end{aligned}
    \end{equation}
where $z_{v,u}$ is the number of triangles that includes the edge $e(v,u)$ in network. $k_{v}$ and $k_{u}$ are the degrees of node $u$ and node $v$, respectively.
    \begin{equation}\label{adjmatix}
    \begin{aligned}
    \begin{split}
    SoECC(v)=\sum_{u\in N_{v}}ECC(v,u)
    \end{split}
    \end{aligned}
    \end{equation}
where $N_{v}$ denotes the set of all neighbors of node $v$.
\item United complex centrality ($UC$) [23]
$$UC(v)=\sum_{u\in N_{v}}(\frac{f_{u}+1}{f_{M}+1}\times ECC_{v,u})$$
where $f_{u}$ denotes the frequency of protein $u$ appeared in the known protein complexes, $f_{M}$ is the maximum frequency that a protein appeared in the known protein complexes.
\end{itemize}

Specifically, we compare $CDC$ with other eight previous measures in YDIP and YMIPS networks, and compare $CIBD$ with other eight previous measures using YMIPS and YMBD networks. Step one, we sort proteins from high to low order on the basis of their values of $CDC$, $CIBD$ and other eight previous measures. Step two, we choose the top 100, 200, 300, 400, 500, and 600 proteins as predictive essential proteins, then compare them with the known essential proteins. Finally, we can get the quantity of true essential proteins among these predictive essential proteins. The experimental results of these measures are shown in Figs. 2-4.

From Fig. 2, the quantity of true essential proteins judged by $CDC$ are 79, 152, 221, 272, 316 and 364 from the top 100 to the top 600, respectively, being the best among the seven methods in YDIP network. Besides $CDC$ method, the method of $LBCC$ also has well performance with 74, 135, 204, 261, 307 and 360 essential proteins correctly identified at the same level. By comparison, the true essential proteins determined by $CDC$ method are increased by 5, 17, 17, 11, 9 and 4, respectively. Compared with other recent methods $SoECC$ and $UC$, $CDC$ also performs an excellent improvement. Moreover, the quantity of essential proteins are much more than previous method including $BC$, $SC$ and $EC$. Although $LAC$ has a good performance, our proposed $CDC$ also has better results than it.

From Fig. 3, we can see that $CIBD$ and $CDC$ both perform better than $DC$, $BC$, $SC$, $LAC$, $EC$, $SoECC$ and $UC$ in YMIPS network, except for $LBCC$. The method of $LBCC$ produces the best results at the top of 200, 500 and 600. $CIBD$ performs the same as $LBCC$ at the top of 100 and 300. At the top of 400, the performance of $CDC$ and $CIBD$ are both better than $LBCC$.

From Fig. 4, $CIBD$ performs closely to the $LBCC$ which gains the best performance at top 100, 200, 400 and 600. $CIBD$ attains the best performance at the top of 300 and 500. We can also see these classical methods ($DC$, $BC$, $SC$, $EC$) perform not well in YMBD network. Hence, our new methods $CDC$ and $CIBD$ can determine much more true essential proteins in most cases.

\subsection{Evaluation of six statistical methods and the precision-recall curves}
To further judge these two indicators of $CDC$, $CIBD$ as well as other eight identification measures, the six statistical methods mentioned in the Section Assessment methods are used. From the formulas, we can obtain some more profound meaning. The sensitivity ($SN$) measures the recognition ability of classifiers to identify correct essential proteins, the larger the value is, the better the classifier is. The specificity ($SP$) measures the recognition ability of classifiers to identify correct non-essential proteins. F-measures ($F$) stands for the harmonic mean of precision and sensitivity. The higher the accuracy ($ACC$) is, the better the classifier is. In conclusion, the values for these six statistical method can reflect the quality of indicators.

Hence, we sort proteins from high to low order on the basis of their values of these methods; Then we take the top 20 percent proteins into account as predictive essential proteins, the remaining 80 percent can be considered as candidates for nonessential proteins. Compared with the known essential protein dataset, we can obtain the values of $TP$, $TN$, $FP$ and $FN$. According to the formulas, the values of these six statistical method would be calculated. On the three different networks, the comparisons among the values of $CDC$, $CIBD$ and other eight measures are executed, showing in Table 3.

For YDIP network, these six statistic values for $CDC$ are higher than other previous measures, which show that $CDC$ has a better prediction accuracy. And the values of $BC$ is the lowest, indicating it has poor performance. For YMIPS and YMBD networks, these six statistic values determined by $CIBD$ are similar to $LBCC$ which also has the ability to predict essential proteins accurately.

\begin{table}[htbp]
\centering
\small
\caption{Comparison the results of sensitivity($SN$), specificity($SP$), positive predictive value($PPV$), negative predictive value($NPV$), F-measure($F$) and accuracy($ACC$) of $CDC$, $CIBD$ and other eight previous algorithms.}
\setlength{\tabcolsep}{3pt}
\begin{tabular}{|c|c|c|c|c|c|c|c|}
\hline

\multicolumn{0}{|c|} {Dataset}& Methods & SN & SP & PPV & NPV & F & ACC \\\hline
\multirow{8}*{YDIP} & DC & 0.363 & 0.825 & 0.416 & 0.789& 0.388 & 0.706\\
\cline{2-8}         & BC & 0.281 & 0.798 & 0.354 & 0.738& 0.313 &0.652\\
\cline{2-8}         &LAC &0.408 & 0.839 & 0.467 & 0.804 & 0.435 & 0.729\\
\cline{2-8}         &SC & 0.335 & 0.811 & 0.36& 0.794 & 0.347 &0.697\\
\cline{2-8}         &LBCC &0.436 &0.853 & 0.512& 0.817&0.477 &0.749\\
\cline{2-8}         &EC & 0.344& 0.814& 0.370 & 0.796 & 0.356 &0.701\\
\cline{2-8}         &SoECC &0.40 &0.850 & 0.463& 0.813&0.428 &0.739\\
\cline{2-8}         &UC &0.391 &0.850 & 0.458& 0.811&0.422 &0.737\\
\cline{2-8}         &CDC &$\mathbf{0.448}$&$\mathbf{0.868} $&$\mathbf{0.515}$ &$\mathbf{0.835}$&$\mathbf{0.487}$ & $\mathbf{0.764}$\\
 \cline{1-8}
\multirow{8}*{YMIPS} & DC &0.274&0.821& 0.305 &0.797 &0.289&0.699\\
\cline{2-8}         & BC &0.197&0.796 &0.278&0.716&0.231 &0.629\\
\cline{2-8}         &LAC &0.287 & 0.825& 0.321& 0.801& 0.303& 0.705\\
\cline{2-8}         &SC &0.139 & 0.782& 0.155 & 0.759 &0.146& 0.638\\
\cline{2-8}         &LBCC &0.430&0.866 & 0.480&$\mathbf{0.841}$ & $\mathbf{0.454}$  & $\mathbf{0.769}$\\
\cline{2-8}         &EC &0.123&0.774&0.155&0.723&0.137&0.610\\
\cline{2-8}         &SoECC &0.281&0.814 & 0.325&$ 0.781$ & $ 0.302$&0.686\\
\cline{2-8}         &UC&0.271 & 0.812 & 0.314 &0.778 & 0.291& 0.682\\
\cline{2-8}         &CDC&0.376 &$\mathbf{0.868}$&$\mathbf{0.530}$ &0.780 & 0.421& 0.723\\
\cline{2-8}         &CIBD &$\mathbf{0.461}$&0.862&$\mathbf{0.503}$&0.778&0.421&0.723\\
 \cline{1-8}

\multirow{7}*{YMBD} & DC &0.261&0.868&0.438&0.749&0.327&0.696\\
\cline{2-8}         & BC &0.244&0.861&0.408&0.743&0.305&0.686\\
\cline{2-8}         &LAC &0.247&0.862&0.413&0.744&0.309&0.688\\
\cline{2-8}         &SC &0.191&0.840&0.320&0.724&0.239&0.657\\
\cline{2-8}         &LBCC &$\mathbf{0.373}$&$\mathbf{0.910}$&$\mathbf{0.617}$&$\mathbf{0.789}$&$\mathbf{0.465}$&$\mathbf{0.760}$\\
\cline{2-8}         &EC &0.219&0.851&0.366&0.734&0.274&0.672\\
\cline{2-8}         &SoECC &0.266 &0.835 & 0.422&0.715&0.326&0.657\\
\cline{2-8}         &UC &0.274 &0.838& 0.434&0.718&0.336&0.662\\
\cline{2-8}         &CIBD &$\mathbf{0.347}$&$\mathbf{0.910}$&$\mathbf{0.581}$&$\mathbf{0.777}$&$\mathbf{0.434}$&$\mathbf{0.745}$\\
\hline
\end{tabular}
\end{table}

\begin{figure}[htbp]
\centering
\includegraphics[width=4in]{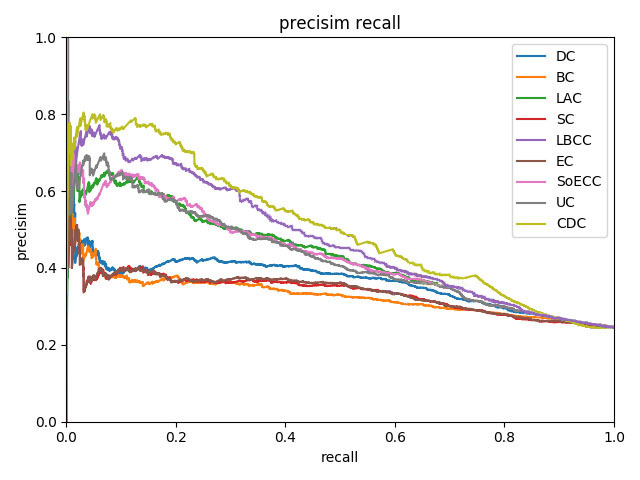}
\scriptsize\\{Fig. 5~Precision and recall curves of $CDC$ and other eight methods for YDIP network.}
\centering
\end{figure}

\begin{figure}[htbp]
\centering
\includegraphics[width=4in]{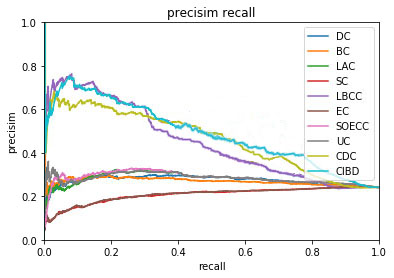}
\scriptsize\\{Fig. 6~Precision and recall curves of $CDC$, $CIBD$ and other eight methods for YMIPS network.}
\centering
\end{figure}
\begin{figure}[htbp]
\centering
\includegraphics[width=4in]{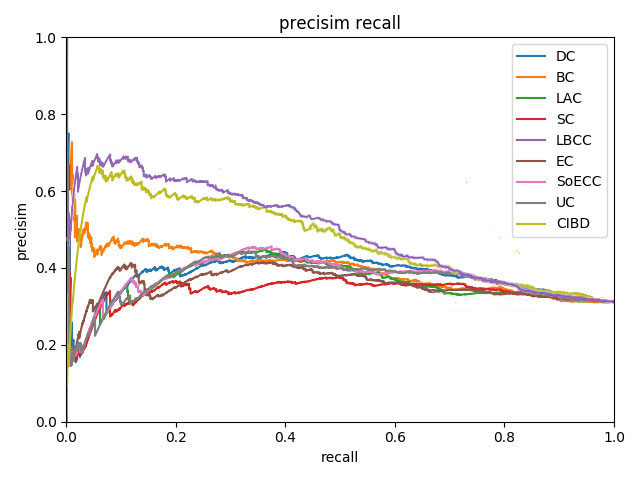}
\scriptsize\\{Fig. 7~Precision and recall curves of $CIBD$ and other eight methods for YMBD network.}
\centering
\end{figure}

In addition, the Precision-Recall curve, a statistical method for evaluating stability, can be used for $CDC$ and $CIBD$ methods and other previous eight measures which defined as follows:
$$Precision(n)=\frac{TP(n)}{TP(n)+FP(n)}$$
$$Recall(n)=\frac{TP(n)}{TP(n)+FN(n)}$$
where the definitions of $TP$, $FP$, $FN$ are depicted in the Assessment method Section. The results are revealed in Figs. 5-7. In YDIP network, our method of $CDC$ has better performance than the other methods. In YMIPS and YDIP networks, the performance of $CDC$ and $CIBD$ are similar to the performance of $LBCC$.

\subsection{Evaluation of jackknife methodology}
Holman et al. developed the jackknife methodology which is an effective universal prediction method [32]. The X-axis represents the quantity of selected predictive essential proteins after sequencing, and the Y-axis represents the quantity of true essential proteins in the selected proteins. The area under the curve reflects the performance of each method. The larger the area under the curve is, the better the centrality is.

First, according to the predicted value, proteins are sorted in descending order. And then we choose predictive essential proteins of top 600 for each dataset. Last, the jackknife curve is drawn based on the accumulation quantity of real essential proteins.

\begin{figure}[htbp]
\centering

\subfigure{
\begin{minipage}[t]{1\linewidth}
\centering
\includegraphics[width=4in]{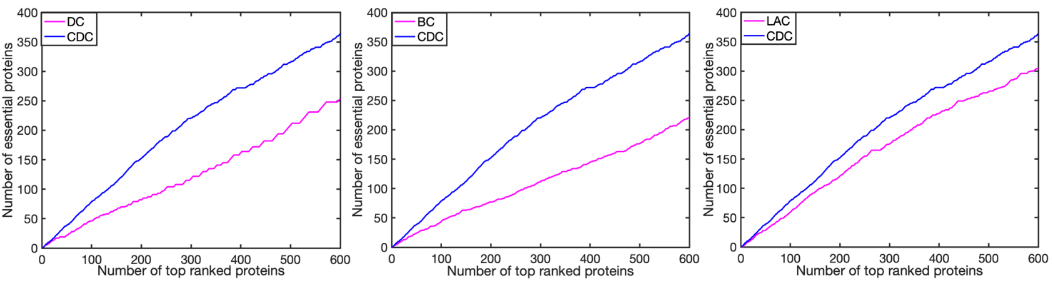}
\end{minipage}%
}%

\subfigure{
\begin{minipage}[t]{1\linewidth}
\centering
\includegraphics[width=4in]{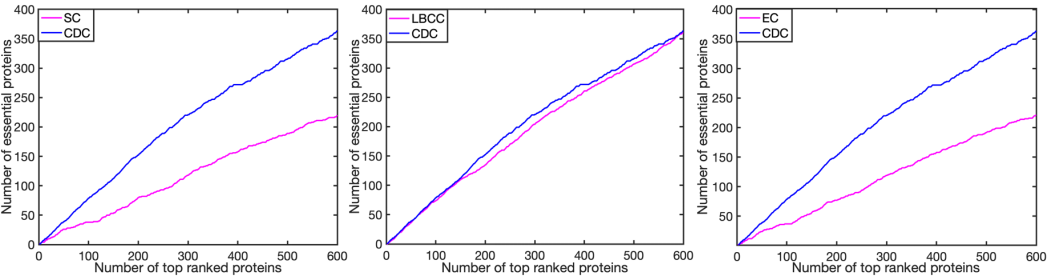}
\end{minipage}
}
\subfigure{
\begin{minipage}[t]{1\linewidth}
\centering
\includegraphics[width=3in]{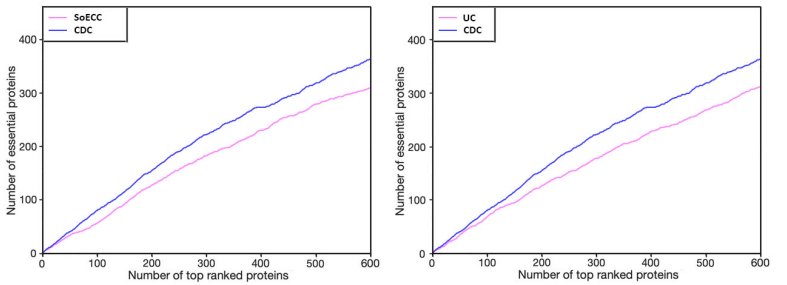}
\end{minipage}
}

\scriptsize
{Fig. 8 The performances of $CDC$ and other eight centrality measures on the YDIP network are evaluated by a jackknife methodology.}
\centering
\end{figure}

\begin{figure}[htbp]
\centering

\subfigure{
\begin{minipage}[t]{1\linewidth}
\centering
\includegraphics[width=4in]{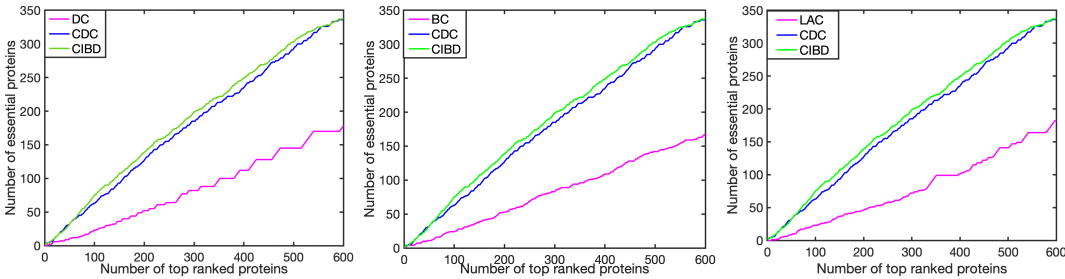}
\end{minipage}%
}%

\subfigure{
\begin{minipage}[t]{1\linewidth}
\centering
\includegraphics[width=4in]{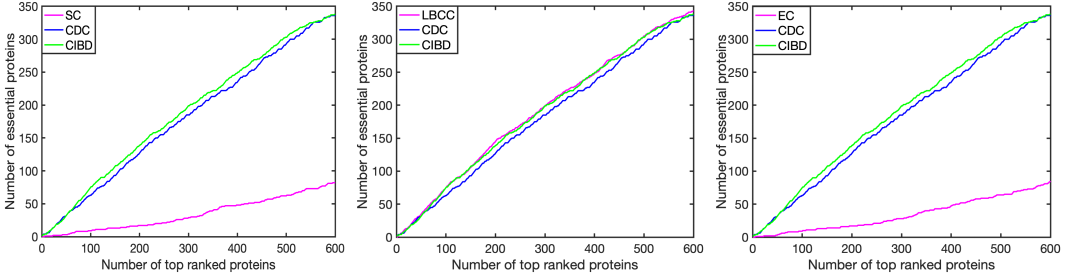}
\end{minipage}
}
\subfigure{
\begin{minipage}[t]{1\linewidth}
\centering
\includegraphics[width=3in]{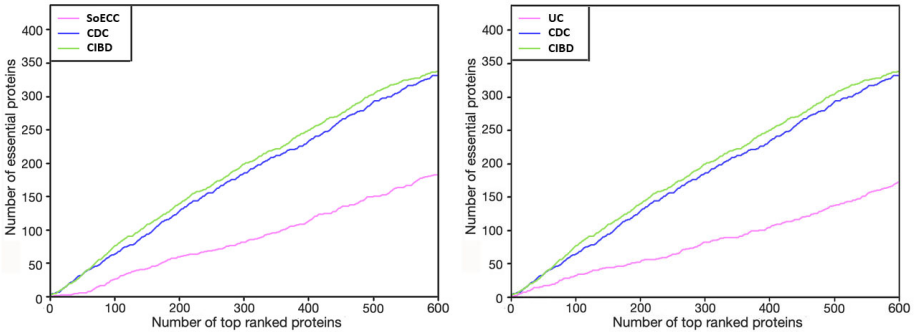}
\end{minipage}
}
\scriptsize
{Fig. 9 The performances of $CDC$, $CIBD$ and other eight centrality measures on the YMIPS network are evaluated by a jackknife methodology.}
\centering
\end{figure}

\begin{figure}[htbp]
\centering

\subfigure{
\begin{minipage}[t]{1\linewidth}
\centering
\includegraphics[width=4in]{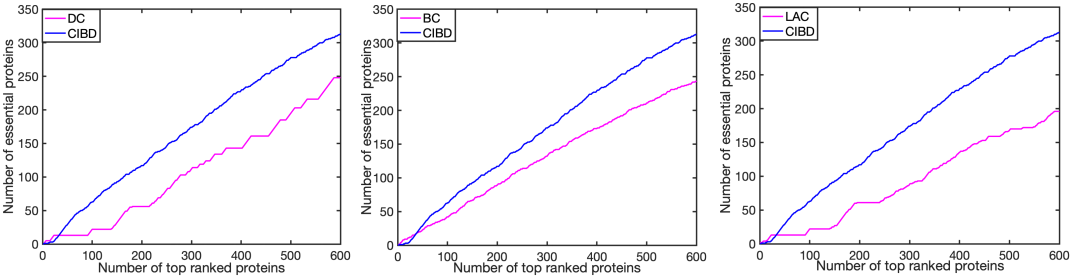}
\end{minipage}%
}%

\subfigure{
\begin{minipage}[t]{1\linewidth}
\centering
\includegraphics[width=4in]{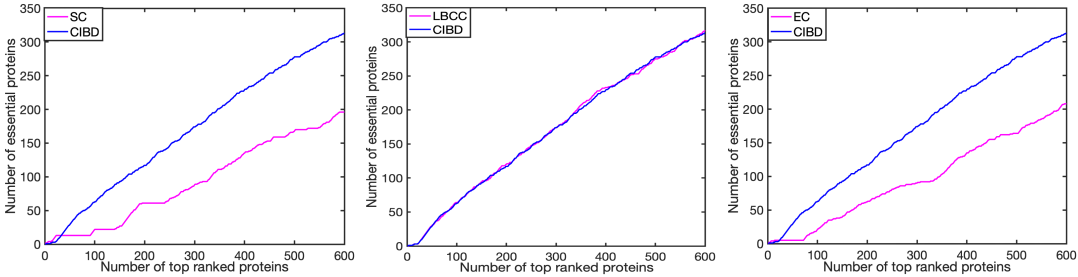}
\end{minipage}
}
\subfigure{
\begin{minipage}[t]{1\linewidth}
\centering
\includegraphics[width=3in]{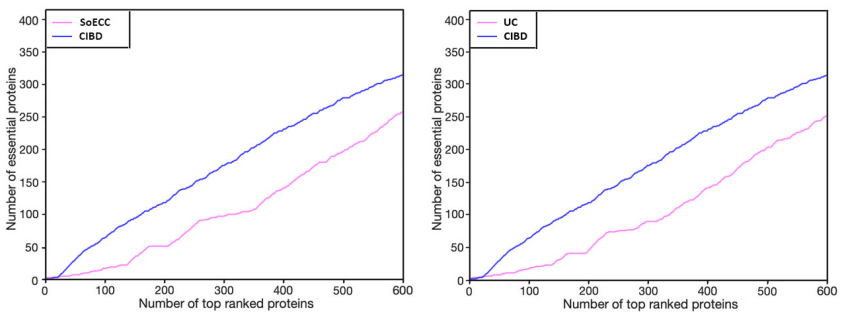}
\end{minipage}
}

\scriptsize
{Fig. 10 The performances of $CIBD$ and other eight centrality measures on the YMBD network are evaluated by a jackknife methodology.}
\centering
\end{figure}

From Fig. 8, it can be seen that the prediction efficiency of $CDC$ is higher than that of other centrality measures on the YDIP network. From Fig. 9, it is shown that $CDC$ and $CIBD$ exhibit performances resemble to that of $LBCC$ and better than those of all the other methods including $DC$, $BC$, $LAC$, $SC$ and $EC$, $SoECC$ and $UC$ on the YMIPS network. From the YMBD network, we can get the same results as shown in Fig. 10. Consequently, the jackknife curves reveal that our methods $CDC$ and $CIBD$ both are effective approaches for predicting essential proteins.

\section{Conclusion}
Identifying essential proteins in protein networks is an indispensable point in the post-genomic era. Improving the recognition rate of essential proteins is a challenging task. At present, plenty of centrality algorithms have been proposed to determine the essentiality of proteins, most of them focus on the analysis and mining of node topology characteristics. In this paper, on the basis of the combination of the local features of protein complexes and topological properties, two new methods are proposed which named as $CDC$ and $CIBD$. We apply them to different datasets YDIP, YMIPS and YMBD. Then we compare the quantity of true essential proteins predicted by $CDC$, $CIBD$ and other eight proposed methods, containing $DC$, $BC$, $LAC$, $SC$, $LBCC$, $EC$, $SoECC$ and $UC$. The results show that $CDC$ and $CIBD$ perform well in most cases. By using the methods of the six statistical, the precision-recall curve and jackknife, we can find that our proposed methods of $CDC$ and $CIBD$ have the ability to improve the accuracy in predicting essential proteins. In future work, deepening the mining of protein biological function and biological significance can be another direction to find the essential proteins.

\end{document}